\documentclass[10pt,conference]{IEEEtran}
\usepackage{graphicx,times,amsmath,amsfonts}
\usepackage{listings}
\usepackage{color}
\usepackage{amssymb}
\usepackage{algorithm}
\usepackage{algorithmic}

\title{A Customized Lattice Reduction Multiprocessor for MIMO Detection}
\author{
\IEEEauthorblockN{Shahriar Shahabuddin, Janne Janhunen, \\Zaheer Khan, and Markku Juntti}
\IEEEauthorblockA{Centre for Wireless Communications\\
University of Oulu, Finland\\
Email: firstname.lastname@ee.oulu.fi \vspace{-1 cm}}\\

\and
\IEEEauthorblockN{Amanullah Ghazi}
\IEEEauthorblockA{Department of Computer Science and Engineering\\
University of Oulu, Finland\\
Email: firstname.lastname@ee.oulu.fi \vspace{-1 cm}}

}

\newcommand{\MYheader}{\smash{\scriptsize
\hfil\parbox[t][\height][t]{\textwidth}{\centering {\normalsize
Reference Matlab code is available at $ sites.google.com/site/shahriarshahabuddin/matlab\_simulator $}}\hfil\hbox{}}}
\makeatletter
\def\ps@headings{%
\def\@oddhead{\MYheader}
\def\@evenhead{\MYheader}
\def\@oddfoot{  }%
\def\@evenfoot{  }}
\def\ps@IEEEtitlepagestyle{%
\def\@oddhead{\MYheader}%
\def\@evenhead{\MYheader}%
\def\@oddfoot{   }%
\def\@evenfoot{   }}
\makeatother
\begin{document}    
\maketitle

\begin{abstract}
\boldmath{Lattice reduction (LR) is a preprocessing technique for multiple-input multiple-output (MIMO) symbol detection to achieve better bit error-rate (BER) performance. In this paper, we propose a customized homogeneous multiprocessor for LR. The processor cores are based on transport triggered architecture (TTA). We propose some modification of the popular LR algorithm, Lenstra-Lenstra-Lov\'{a}sz (LLL) for high throughput. The TTA cores are programmed with high level language. Each TTA core consists of several special function units to accelerate the program code. The multiprocessor takes 187 cycles to reduce a single matrix for LR. The architecture is synthesized on 90 nm technology and takes 405 kgates at 210 MHz.
 }
\end{abstract}
\section{Introduction}\label{1}

Multiple-input multiple-output (MIMO) is a key technology to utilize the available radio spectrum efficiently. The basic idea of MIMO is to send multiple independent data streams from multiple antennas in the same frequency band. These independent streams need to be separated at the receiver to identify the symbol that is being transmitted by using a MIMO detector. Maximum likelihood (ML) is the optimal solution for the MIMO detection problem that compares the incoming symbol with every possible symbols in the constellation. However, the ML algorithm is too complex for practical real-time implementations. Linear detection is popular for practical implementations. The linear MIMO detection algorithms are less complex, but suffers from a degraded bit error-rate (BER) performance.

Lattice reduction (LR) is a preprocessing technique that can be used with the linear detection to significantly improve the BER performance and reduce the gap between the traditional linear detectors and optimal ML. LR transforms the MIMO channel matrix to a near orthogonal matrix and thus facilitates to achieve a better BER performance. The most used LR algorithm is called the Lenstra-Lenstra-Lov\'{a}sz (LLL) algorithm according to the name of the inventors [\ref{LLL}]. The LLL algorithm poses many challenges due to the undeterministic execution time and higher computational complexity. We propose a modified LLL (MLLL) algorithm that is based on the original LLL algorithm on complex domain. We use a fixed structure for the LLL based on [\ref{fixed}]. Instead of using the Lov\'{a}sz condition, a less complex Siegel condition is applied [\ref{seysen}]. An early termination technique is used as proposed in [\ref{bruderer}]. We demonstrate by Matlab simulation that the BER performance loss of the hardware friendly MLLL algorithm is negligible.

There are several hardware accelerators proposed in [\ref{bruderer}] [\ref{clarkson}] [\ref{shabany}] [\ref{tcas2}]  for different LR algorithms. The fixed hardware implementations provide high data rate and consume less silicon area compared to the customized application specific processors (ASIP). The drawback of the fixed hardware implement
ation is that it operates only on a fixed set of parameters due to the hardwired control path and it is not possible to modify the control path in the future. An ASIP customized for a small set of algorithms is an attractive solution in terms of cost, silicon area and high throughput. Most importantly, an ASIP reduces the design risk with an instruction memory that can be used to load new programs or control instructions. The control instructions can be easily obtained by a retargetable compiler for that particular customized architecture.

%
%

A customized very long instruction word (VLIW) processor is implemented in [\ref{ubaid}] for the LR. We take different approach and design a customized multiprocessor based on the transport triggered architecture (TTA) paradigm. TTA is a processor design philosophy where the programmer can control the internal data transports between different function units of the processor. TTA exploits the instruction level parallelism (ILP) by processing several instructions in a single clock cycle. The TTA based codesign environment (TCE) tool is used in this work to design the TTA processor cores. TCE enables the designer to write an application with a high level language and design the target processor in a graphical user interface at the same time. A turbo decoder and a MIMO detector design using TCE can be found in [\ref{vecturbo}] and [\ref{mutimode}]. In this work, every core of the proposed multiprocessor is programmed with C language to shorten the time-to-market. The multiprocessor takes 187 cycles and achieves a maximum clock frequency of 210 MHz on 90 nm technology. To our knowledge, this is the first TTA based customized architecture for LR.


\section{System model} \label{2}

\subsection{Conventional MIMO Detection}

Consider a MIMO system consists of $\it{M_T}$ transmit antennas, which are sending data over the channel and $\it{N_R}$ receive antennas which are receiving transmitted bits from the channel. The modulation scheme that is used here is quadrature amplitude modulation (QAM) with the assumption $ N_R \geq M_T $. The received signal $\mathbf{y}$ can be represented as

\begin{equation}\label{gf}
    \mathbf{y}= \mathbf{H} \mathbf{x}+\mathbf{n},
\end{equation}
where $\mathbf{y}\in \mathbb{C}^{N_R}$ is the received signal vector, $\mathbf{x} \in \mathbb{C}^{M_T}$ is the transmit symbol vector, $\mathbf{H} \in\mathbb{C}^{N_R \times M_T}$ is the channel matrix and $\mathbf{n} \in \mathbb{C}^{N_R}$ is the circularly symmetric complex white Gaussian noise vector with zero mean and variance $\sigma^2$.

In the receiver, the linear zero forcing (ZF) detector calculates the inverse of the channel matrix to compute the transmitted symbol vector which can be expressed by,

\begin{equation}
    \mathbf{\tilde{x}=(H^\text{H}H)^{\text{-1}}Hx}=\mathbf{H^\dag x}.
\end{equation}
where $\mathbf{H}$ is the channel matrix and $ (\cdot)^\dag$ denotes the pseudoinverse.  Typically, the channel matrix $\mathbf{H}$ is QR decomposed into two parts as $\mathbf{H}=\mathbf{Q}\mathbf{R}$.
Here $\mathbf{Q}\in \mathbb{C}^{(N_R \times N_R)}$ denotes a unitary matrix and $\mathbf{R} \in \mathbb{C}^{(N_R \times N_R)}$ denotes an upper triangular matrix.

\subsection{Lattice Reduction}

A lattice is a periodic arrangement of discrete points. A lattice can be characterized in terms of a set
of basis vectors, where any points of the lattice can be represented by a superposition of integer multiples
of the basis vectors. For simplicity, we call the set $\mathbf{B} = (b_1,b_2,....,b_n)$ as the basis of the lattice.

A complex valued lattice in the $n$-dimensional complex space $\mathbb{C}^n$ can be defined as

\begin{equation}
    \mathcal{L}=\{\boldsymbol{\upsilon}|\boldsymbol{\upsilon} = \mathbf{B}\boldsymbol{\omega}\},
\end{equation}
where $\mathbf{B}$ is the basis of the lattice and $\boldsymbol{\omega} = [\omega_1,\omega_2,....,\omega_n]$.
Note that in (3), the $\boldsymbol{\upsilon}$, $\boldsymbol{\omega}$ and matrix $\mathbf{B}$  can be replaced with  $\mathbf{y}$, $\mathbf{x}$ and $\mathbf{H}$ respectively to obtain
    $\mathcal{L}=\{\mathbf{y} | \mathbf{y}  = \mathbf{Hx}\}$. In this case, the vector space $\mathcal{L}$ is the set of all possible undisturbed received signal
points. There are many bases that can span the space $\mathcal{L}$, and the aim of the LR algorithm is to find a set of least correlated base with the shortest basis vectors [\ref{magazine}].

\subsection{LR-based MIMO Detection}

LR finds an improved basis for the lattice induced by the channel. The original basis and the reduced basis are related by a unimodular matrix, $\mathbf{T}$. Therefore, the LR aided detection finds the received symbol in the new reduced basis and afterwards transfer the signal in the original lattice. The new channel matrix after the LR can be expressed as, $\tilde{\mathbf{H}} =  \mathbf{H}\mathbf{T}$ and the transmitted
signal is also treated as multiplied by $\mathbf{T}^{-1}$ which is $\mathbf{z} = \mathbf{T}^{-1}\mathbf{x}$ for the reduced basis.
The received signal $\mathbf{y}= \mathbf{H} \mathbf{x}+\mathbf{n}$ can be expressed as
\begin{equation}
\mathbf{y} = \mathbf{HT}\mathbf{T}^{-1}\mathbf{x} + \mathbf{n} = \tilde{\mathbf{H}}\mathbf{z} + \mathbf{n}.
\end{equation}
The LR aided detection operates on $\tilde{\mathbf{H}}$ and $\mathbf{z}$ instead of $\mathbf{H}$ and $\mathbf{x}$. The LR aided ZF detector can be expressed as
\begin{equation}
    \mathbf{\tilde{\mathbf{x}}=(\tilde{H}^\text{H}\tilde{H})^{\text{-1}}\tilde{H}z}=\mathbf{\tilde{H}^\dag z}.
\end{equation}
The LR algorithm is applied on the QR decomposed $\mathbf{H}$ to obtain the modified  $\tilde{\mathbf{Q}}$ and  $\tilde{\mathbf{R}}$. Afterwards, the lattice reduced channel matrix can be obtained as $\tilde{\mathbf{H}} = \tilde{\mathbf{Q}}\tilde{\mathbf{R}}$.

\section{Lattice Reduction Algorithm} \label{3}

LLL algorithm is widely used to compute the suitable unimodular matrix $\mathbf{T}$ and to obtain a reduced lattice basis. LLL was originally proposed for the real valued LR [\ref{LLL}]. However, the channel matrix is naturally complex valued and therefore, complex version of LLL (CLLL) is used to reduce the complexity.

The CLLL algorithm suffers from irregular dataflow, which eventually leads to higher latency. Therefore,  a fixed-complexity LLL (fcLLL) algorithm is proposed in [\ref{fixed}]. The fcLLL alters the signal flow of the CLLL to follow a deterministic structure. It is possible to utilize less complex Siegel algorithm instead of the complex Lov\'{a}sz condition [\ref{seysen}]. It is also very important to use an early termination mechanism to meet the strict requirements. Applying all this modifications, we propose a modified-LLL (MLLL) algorithm for LR with less complexity and negligible BER performance loss. The MLLL implemented in this paper is summarized in Algorithm 1.

\begin{algorithm}
\caption{\textbf{Modified CLLL Algorithm (MLLL)}}
\label{algo1}
\begin{algorithmic}
\STATE{\text{INPUT}: $\mathbf{Q}\in\mathbb{C}^{N_R\times N_R}$ , $\mathbf{R}\in\mathbb{C}^{N_R\times N_R}$ , $\delta$  }
\STATE{1: \text{Initialization} $\mathbf{\tilde{Q}}:=\mathbf{Q}$ , $\mathbf{\tilde{R}}:=\mathbf{R}$ , $\mathbf{T}:=\mathbf{I}_{M_T}$ }
\STATE{2: $k := 2 $}
\STATE{3: \hspace{2 mm} \textbf{while}  $k \leq iterations$ }
\STATE{4: \hspace{4 mm} \textbf{for} $l = k - 1$ \textbf{to} $1$ \textbf{step} $-1$ }
\STATE{5: \hspace{8 mm} $\mu = \mathbf{\tilde{R}}(l,k)/\mathbf{\tilde{R}}(l,l)$  }
\STATE{6: \hspace{8 mm} \textbf{if} $\mu \neq 0$ }
\STATE{7: \hspace{10 mm} $\mathbf{\tilde{R}}(1:l,k) := \mathbf{\tilde{R}}(1:l,k) - \mu \mathbf{\tilde{R}}(1:l,l) $}
\STATE{8: \hspace{10 mm} $\mathbf{T}(:,k) := \mathbf{T}(:,k) - \mu \mathbf{T}(:,l)$  }
\STATE{9: \hspace{8 mm} \textbf{end} }
\STATE{10: \hspace{4 mm} \textbf{end}  }
\STATE{11: \hspace{4 mm} \textbf{if} $\delta \mathbf{\tilde{R}}(k-1,k-1)^2 > \mathbf{\tilde{R}}(k,k)^2  $ }
\STATE{12: \hspace{8 mm} Swap columns $k-1$ and $k$ in $\mathbf{\tilde{R}}$ and $\mathbf{T}$ }
\STATE{13: \hspace{8 mm} $\Theta =  \begin{bmatrix} \alpha & \beta \\[0.3em]  -\beta & \alpha \end{bmatrix}$ with $ \alpha = \frac{\mathbf{\tilde{R}}(k-1,k-1)}{\|\mathbf{\tilde{R}}(k-1:k,k-1)\|} $ and $\beta = \frac{\mathbf{\tilde{R}}(k,k-1)}{\|\mathbf{\tilde{R}}(k-1:k,k-1)\|} $}
\STATE{14: \hspace{8 mm} $\mathbf{\tilde{R}}(k-1:k,k-1:k) := \Theta\mathbf{\tilde{R}}(k-1:k,k-1:k) $    }
\STATE{15: \hspace{8 mm} $\mathbf{\tilde{Q}}(:,k-1:k) := \mathbf{\tilde{Q}}(:,k-1:k)\Theta^T $ }
\STATE{16: \hspace{8 mm} $k:= \text{max}\{k-1,2\}$  }
\STATE{17: \hspace{4 mm} \textbf{else}  }
\STATE{18: \hspace{8 mm} $k:=k+1$  }
\STATE{19: \hspace{4 mm} \textbf{end}  }
\STATE{20: \hspace{2 mm} \textbf{end}   }

\end{algorithmic}
\end{algorithm}

The BER performance of the traditional ZF, original CLLL aided ZF, MLLL aided ZF and the optimal ML is simulated for various signal-to-noise (SNR) in a Matlab simulator. An additive white Gaussian noise (AWGN) channel is used for 16-QAM modulation and the BER is averaged over 10 000 Monte-Carlo trials. Fig. 1 shows the MLLL algorithm with 5 iterations. It can be seen that the performance loss is negligible compared to the original CLLL algorithm.

\begin{figure}[h]
\centering
\includegraphics[keepaspectratio,width=1\columnwidth]{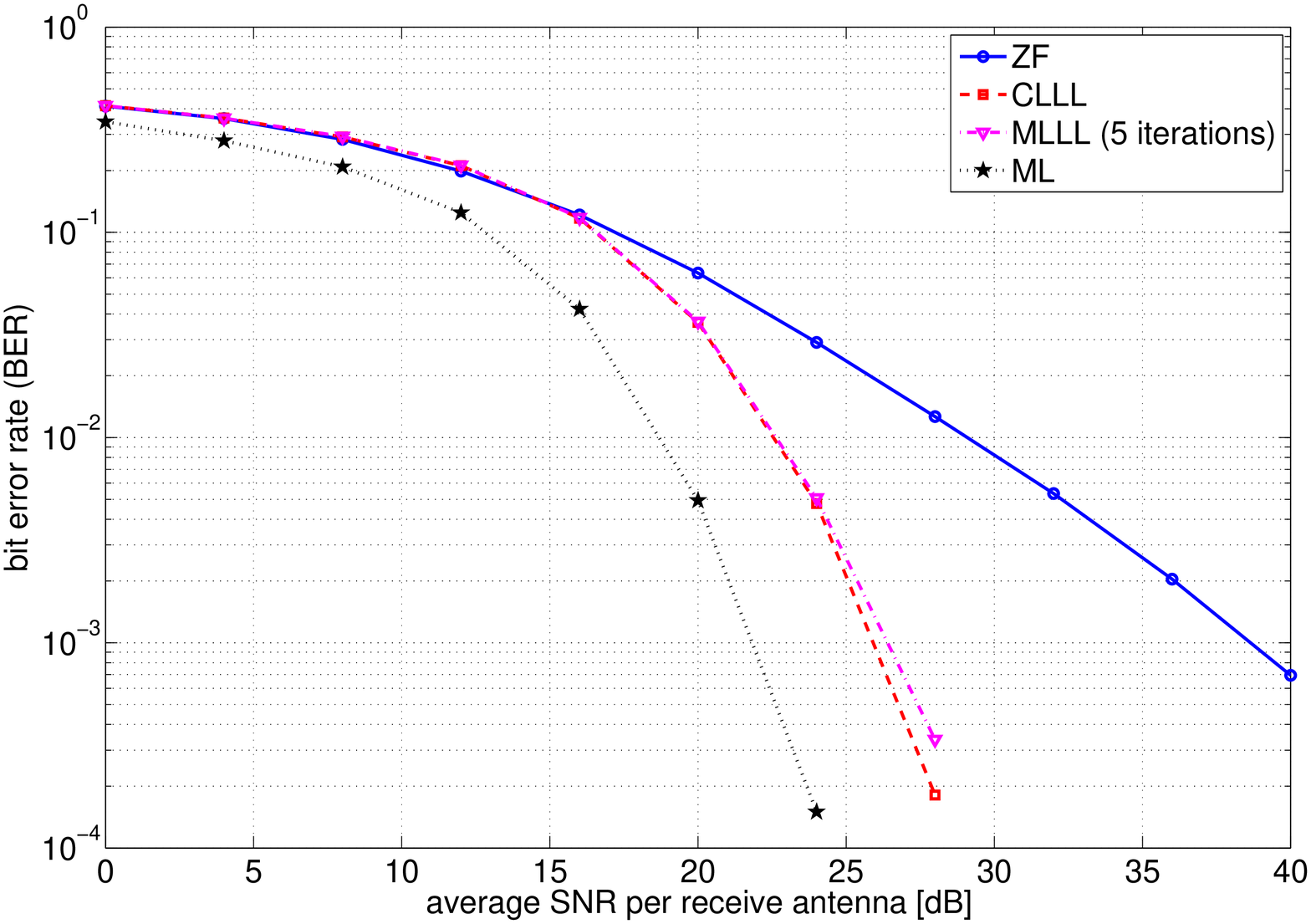}
\caption{BER peformance of MLLL algorithm.}
\label{fig:}
\end{figure}

\section{TTA Multiprocessor for MLLL}\label{4}

\subsection{Special Function Units}

Six special function units (SFU) are designed and written in VHSIC hardware description language (VHDL) to accelerate each iteration of the MLLL algorithm. A special function unit is designed to support the complex multiplication (CMUL) operation. Data level parallelism (DLP) is applied in the design by packing the 16-bit real part and 16-bit complex part in a 32-bit complex variable. Therefore, CMUL uses four 16-bit multipliers, a single 16-bit adder and 16-bit subtractor to support the complex multiplication.

Two SFUs for $\mu$ calculation and size reduction are designed according to [\ref{ubaid}]. These SFUs are single-cycle and multiplier-less. It is observed from the Matlab simulations that the value of $\mu$ has a range of $[-4,4]$, and thus, the dynamic range of the SFUs are set accordingly. Another simple SFU is designed to compute the SIEGEL criterion. Instead of multiplying the input with $.75$ the SIEGEL SFU calculates the value with a combination of two shifters and one adder.

\begin{figure}[h]
\centering
\includegraphics[keepaspectratio,width=.9\columnwidth]{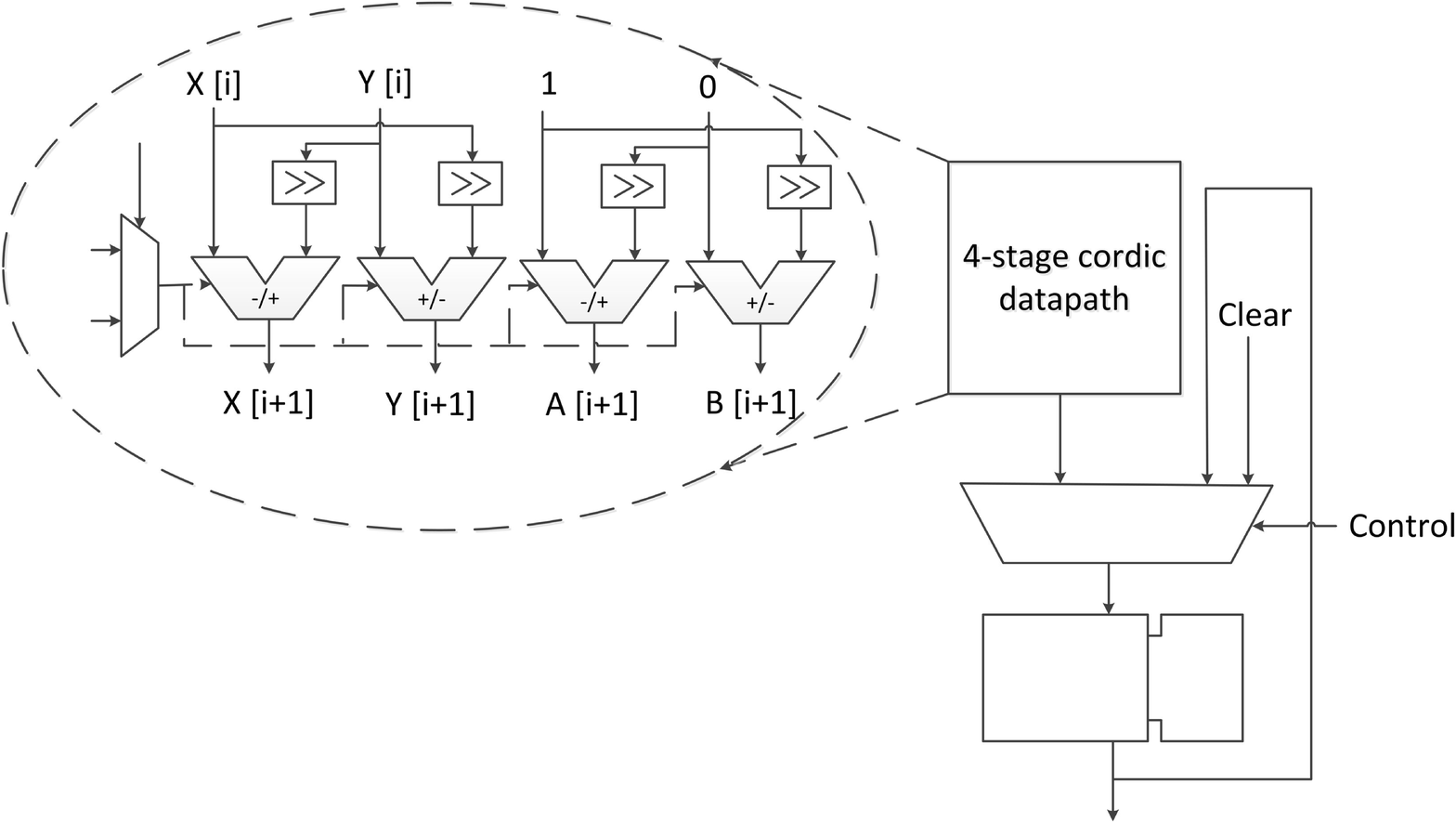}
\caption{4-cycle cordic architecture.}
\label{fig:}
\end{figure}

The most complex SFU that lies in the datapath of a single TTA core is the CORDIC SFU. A master-slave cordic is considered in this work [\ref{bruderer}]. The master-slave CORDIC is a combination of two CORDIC blocks in vectoring mode and rotation mode respectively. By setting the input as 1 and 0 of the CORDIC with rotation mode, it is possible to calculate the cosine and sine values directly. Therefore, the angle calculation done in a conventional CORDIC block is not needed here. In every stage it is possible to calculate the values of the signums and add or subtract accordingly in the rotation mode.
As we need a 16-bit CORDIC, there are two options to design it. An iterative CORDIC that uses registers and iterates 16 times over the 1-stage datapath. However, it takes 16-clock cycles to compute the output. For a processor based implementation a 16-cycle SFU is complex as there will be 15 NOP operations in the assembly code. It is possible to fully unroll the CORDIC block without any registers. Then the critical path for the CORDIC block becomes too high. We find a compromise between the two approach and design a 4-stage CORDIC datapath that can be reused four times to create a 4-cycle master-slave CORDIC. The block diagram of the master-slave CORDIC is presented in Fig. 2, where the ellipse contains a single stage of the datapath. An ARRANGE SFU is designed to rearrange the 32-bit variables.

\subsection{High Level Architecture of the multiprocessor}
A 32-bit fixed point TTA processor is designed to support a single iteration of the MLLL algorithm and five of these TTA cores are connected in a pipelined fashion to compute the LR matrix.
Part of a single TTA processor core is illustrated inside the dotted block of Fig. 3. For readability, the whole processor is not given. The blocks in the upper part of the core represent the function units and register files of the processor. The black horizontal straight lines represent the buses of the processor. The vertical rectangular blocks represent the sockets.


\begin{figure}[h]
\centering
\includegraphics[width=\columnwidth,height=2.8cm]{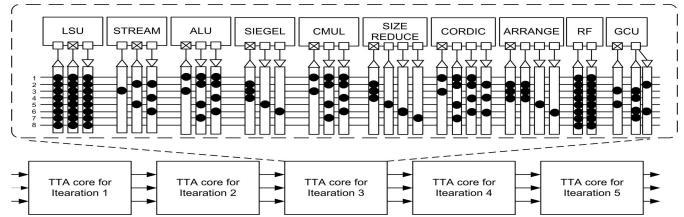}
\caption{The multiprocessor architecture.}
\label{fig:}
\end{figure}

 Each core includes the load/store unit (LSU), arithmetic logic unit (ALU), global control unit (GCU), register files, several conventional function units and SFUs. The $\mathbf{Q}$, $\mathbf{R}$ and $\mathbf{T}$ matrix are read from three separate first-in-first-out (FIFO) memory buffer by using the function units called STREAM. The STREAM units can read every input sample in one clock cycle. Three STREAM units are used to get the inputs simultaneously. Three STREAM unit is used to write the outputs in the memory buffer.

Ten register files are used to save the intermediate results. A single Boolean register file is included in the processor design. When the registers are not enough, the processor is able to access the data memory to temporary store data through the LSU. The SFUs can be called by macros to accelerate the program code.


Eight buses are used in a single core and therefore the core is able to process eight instruction in a single clock cycle. As the datapath is exposed to the programmer in the TTA architecture, it is possible to remove the unused or less frequently used connection between the function units and buses. Thus, several connections of the processor is removed to reduce the cost of a core. The connection between function units and buses is illustrated by black spots in the sockets of Fig. 3.

The cores are connected to one another by FIFO buffer memories. In this way, five iteration of MLLL can be processed in parallel. The cores are identical and except the first core, the rest of the cores use the same program image.

\section{Results and Discussion} \label{5}

It can be seen from the TCE tool that the TTA multiprocessor takes 187 cycles to compute the MLLL algorithm. Some of the operations executed during a single iteration is presented in Table \ref{tab3}. The conventional operations like addition, shifts are not shown in the table.

\begin{table}[h]
\centering
\caption{Number of operations}
\label{tab3}
\begin{tabular}{|c|c|}
\hline
Operation & No. of Ops \\
\hline
ARRANGE & 18 \\
\hline
CORDIC &  9\\
\hline
CMUL &  72 \\
\hline
STREAM & 84\\
\hline
SIEGEL & 3\\
\hline
SIZE REDUCTION & 34\\
\hline
\end{tabular}
\end{table}

The multiprocessor is synthesized using UMC 90 nm standard cell library $(fsd0k\_generic\_core 1d0vtc)$. Synopsys Design Compiler is used to estimate gate count and maximum achievable clock frequency. The operating conditions (temperature, operating voltage, manufacturing process quality) for synthesis are set to default value (TCCOM). The maximum clock frequency achieved during the synthesis for the multiprocessor is 210 MHz. The total gate count of the multiprocessor at 210 MHz is around 405 kgates.

A comparison with different other implementations of LR is presented in Table \ref{tab4}. Two important VLSI architectures for the LR algorithm can be found in [\ref{bruderer}] and [\ref{shabany}] with low latency and area. The authors implemented the reverse-siegel LLL (RS-LLL) and hardware-optimized LLL (HOLL) in [\ref{bruderer}] and [\ref{shabany}] respectively. The VLSI architecture for the Clarkson's algorithm is provided in [\ref{clarkson}]. The architecture provides less throughput than our architecture even after using a hardwired control path. The latency of the VLSI architecture of [\ref{tcas2}] is lowest, but with the price of a very low maximum achievable clock frequency of 37 MHz. Though most of the VLSI implementations take less cycles and area, the architectures suffer from inflexibility, and as a consequence later field updates are not possible.

As different variants of LLL algorithms are proposed in different literatures, a flexible implementation is a necessity. Our customized multiprocessor is an example of such a flexible implementation with moderate latency and cost. It is possible to support different variants of LLL algorithms by updating the instruction memory with new binary program image. The updated binary instructions can be obtained by compiling the other LLL algorithms for our particular architecture with the help of a retargetable compiler.

The programmable VLIW core [\ref{ubaid}] takes less clock cycle and flexible. The implementation consisted of not only LR, but also QR decomposition and detection also. However, it is not clear the amount of area needed only for the LR. The total area is very high compared to the other implementations even at 40 nm technology.

\begin{table}[h]
\centering
\caption{Implementation comparison}
\label{tab4}
\begin{tabular}{|c|c|c|c|c|}
\hline

Reference & Architecture/tech. & area & max-freq. & cycles \\
\hline
[\ref{bruderer}] & .13 $\mu$m & 107 kGE & 333 MHz & 14 \\
\hline
[\ref{clarkson}] & Virtex-II Pro & N/A & 100 MHz & 420 \\
\hline
[\ref{shabany}] & .13 $\mu$m & 125 kGE & 352 MHz & 40 \\
\hline
[\ref{tcas2}] & 90 nm & 200 kGE & 37 MHz & 5 \\
\hline
[\ref{ubaid}] & VLIW (40 nm)  & 6364 kGE & 700 MHz & 21 \\
\hline
Proposed & TTA (90 nm) & 405 kGE & 210 MHz & 187 \\
\hline
\end{tabular}
\end{table}

\section{Conclusion} \label{6}

We propose a modified LLL algorithm for LR. We simulated in Matlab the performance of the algorithm and propose a customized multiprocessor architecture for the MLLL. The cores are programmable with the help of a retargetable compiler. The flexible implementation shows great promise to support later field updates and provides high throughput with a moderate cost.


{}

\end{document}